%% file: paper82g.tex
\documentclass[11pt]{article}

\usepackage{epsfig}
\usepackage{graphics}
\usepackage{color}
 
\newcommand{\SLASH}[2]{\makebox[#2ex][l]{$#1$}/}

\newcommand{\eslash}{\SLASH{\epsilon}{.06}}
\newcommand{\lslash}{\SLASH{l}{-.15}}
\newcommand{\uslash}{\SLASH{u}{.15}}
\newcommand{\kslash}{\SLASH{k}{.15}}
\newcommand{\pslash}{\SLASH{p}{.2}}
\newcommand{\qslash}{\SLASH{q}{.08}}

\newcommand{\be}{\begin{equation}}
\newcommand{\ee}{\end{equation}}
\newcommand{\beeq}{\begin{eqnarray}}
\newcommand{\eeeq}{\end{eqnarray}}

\def\epsbo{\mbox{\boldmath $\epsilon$}}
\def\Deltabo{\mbox{\boldmath $\Delta$}}
\def\ubo{{\bf u}}
\def\kbo{{\bf k}}
\def\lbo{{\bf l}}
\def\rbo{{\bf r}}
\def\pbo{{\bf p}}
\def\bbo{\mbox{\boldmath $b$}}
\def\rb{\mbox{\boldmath $r_1$}}
\def\rbb{\mbox{\boldmath $r_2$}}

\begin{document}

\begin{titlepage}
\hfill
\hspace*{\fill}
\begin{minipage}[t]{4cm}
 DESY--03--010\\
 hep-ph/0301192
\end{minipage}
\vspace*{2.cm}
\begin{center}
  \begin{LARGE}
    {\bf On the Dipole Picture in the Nonforward
Direction}\\
  \end{LARGE}
  \renewcommand{\thefootnote}{\fnsymbol{footnote}}
  \renewcommand{\thefootnote}{\arabic{footnote}}
  \setcounter{footnote}{0} 
  \vspace{2.5cm} 
  \begin{Large} 
    {J.Bartels$^{(a)}$, K.Golec-Biernat$^{(b,a)}$ and K.Peters$^{(a)}$}\\
  \end{Large} 
  \vspace{0.3cm} 
  \textit{$^{(a)}$II.\ Institut f\"ur Theoretische Physik,
    Universit\"at Hamburg,\\ Luruper Chaussee 149, 
    D-22761 Hamburg\footnote{email: 
        bartels@x4u2.desy.de $\quad$ golec@mail.desy.de 
        $\quad$ krisztian.peters@desy.de 
        }\\
{\rm and}\\
$^{(b)}$H. Niewodnicza\'nski Institute of Nuclear Physics\\
Radzikowskiego 152, 31-342 Krak\'ow, Poland
        }  \\

\end{center}
\vspace*{1.5cm}
\begin{abstract}
We calculate, for nonzero momentum transfer, 
the dipole formula for the high energy behaviour of elastic and quasielastic 
scattering of a virtual photon. 
We obtain an expression of the nonforward photon impact factor and 
of the nonforward photon wave function, and we give a physical interpretation.

\end{abstract}
\end{titlepage}
\section{Introduction}
The color dipole picture  \cite{Nikolaev,Mueller} of deep inelastic
scattering at small $x$ 
has turned out to be very useful in describing, at low $Q^2$, the transition 
from pQCD to nonperturbative strong interactions. In a first sequence of 
attempts models for the dipole cross section have been formulated in order 
to describe the interaction between the
quark-antiquark pair and the proton in the 
forward direction \cite{GBW,ForshawForw,Levin,McDermott,BGBK}. Equivalently,
in these models the interaction is integrated over the full region 
of impact parameter $b$. More recently, attention has been given to the 
$t$-dependence of diffractive vector production in deep inelastic scattering.
HERA data show significant differences compared to 
hadron-hadron scattering. For example, in $J/\Psi$
production, the $t$-slope
is smaller, indicating a smaller transverse extension of the scattering
system. Also, shrinkage is considerably smaller, hinting at a quite different 
picture in impact parameter space. A phenomenological analysis in
\cite{Mueller} investigates the beginning of saturation at small impact
parameter and low $Q^2$. It therefore seems natural to understand  
the dipole picture at nonzero momentum transfer, and then to 
search for models for the dipole cross section which depend upon the impact 
parameter $b$.

First calculations of the photon impact factor for real photons in the 
nonforward direction have been presented rather long time ago 
~\cite{Cheng-Wu,Lipatov}. The results have been obtained in momentum space, 
but at that time no attempt has been made to find an interpretation 
in terms of the photon wave function and a dipole cross section. 
More recently, the nonforward photon impact factor to 
an off-shell incoming and real outgoing photon with massless quarks 
has been given in ~\cite{Wusthoff-Ivanov,ForshawNonForw}, 
and in ~\cite{Ryskin,IK} the nonforward diffractive production of a 
vector particle has been analysed. Finally,  
in ~\cite{Gieseke-Qiao} the nonforward diffractive production of a (massive) 
quark-antiquark pair has been studied, but the result has not been 
presented in a form which is convenient for the color dipole picture analysis
mentioned above. 

In this paper we present an analysis of the nonforward photon impact 
factor for all photon helicities, using massive quarks and the 
general kinematic $\gamma^*(Q_+^2) \to \gamma^*(Q_-^2)$,
and we present our result in the form of the color dipole picture:
(photon wave function)$\cdot$(color dipole cross section)$\cdot$(photon wave 
function). We find the form of the (nonforward) photon wave function and of 
the dipole cross section, and we give a physical interpretation in the 
infinite momentum frame. Transforming to impact parameter, the dipole cross 
section will be shown to depend upon the distance between one of the quarks of 
the color dipole and the target. For open quark antiquark production we 
show that the integrated diffractive cross section can be expressed in terms 
of the same nonforward dipole cross section as the elastic 
$\gamma^*\to\gamma^*$ scattering amplitude.           

Our paper will be organized as follows. We first (section 2) calculate, 
in momentum space, the high energy behaviour of the elastic scattering process 
$\gamma^* + \gamma^* \to \gamma^* + \gamma^*$ in lowest order QCD.
The resulting formula leads to the nonforward photon impact factor
which, when transformed to transverse coordinate space, leads to the 
photon wave function and to the nonforward dipole scattering amplitude        
(section 3). The transformation to impact parameter is done in section 4,
and a physical interpretation is given in the infinite momentum frame.      
Section 5 briefly describes the nonforward dipole formula for 
$\gamma^*p$ scattering, and in section 6 we present the formula for 
open quark-antiquark production.    

%%%%%%%%%%%%%%%%%%%%%%%%%%%%%%%%%%%%%%%%%%%%%%%%%%%%%%%%%%%%%%%%%%%%%%%%%%
\section{Nonforward impact factors}

To be definitive, let us consider elastic $\gamma^*\gamma^*$ scattering
at high energies and small (but nonzero) momentum transfer. In order 
to have a non-falling cross section we consider the lowest order diagrams 
with two-gluon exchange. The scattering amplitude takes the form:
\begin{eqnarray}
\label{eq:a}
A(s,\lbo)\,=\,\frac{is}2 \int
\frac{d^2{\bf k }}{(2\pi)^3}\,\frac{\Phi(q,k,l)}{{\bf (k+l)}^2 }
\frac{\Phi(p,-k,-l)}{{\bf (k-l)}^2 } \, .
\end{eqnarray}
The four diagrams which contribute to the photon impact factor $\Phi$
are shown in Fig.1   
%%%%%%%%%%%%%%%%%%%%%%%%%%%%%%%%%%%%%%%%%%%%%%%%%%%%%%%%%%%%%%%%%%%%%%%%%%%%  
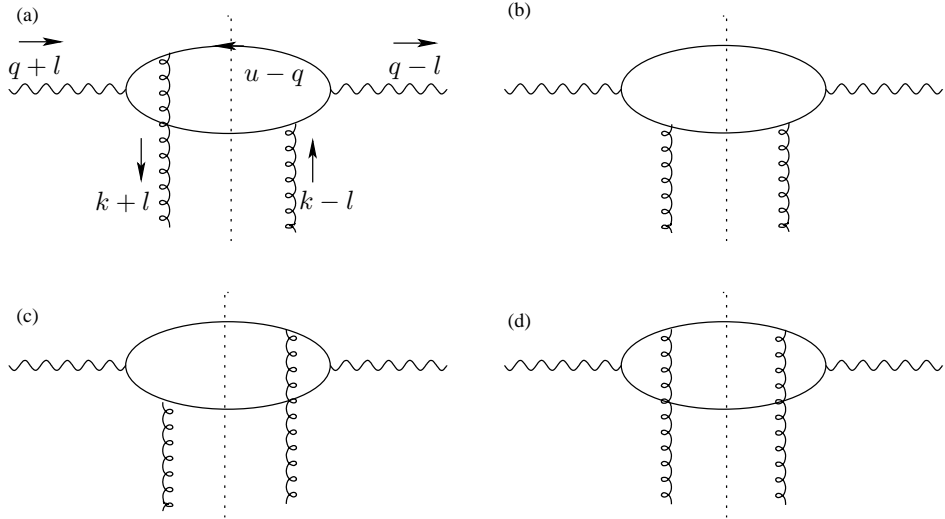
\begin{figure}[t]
\label{fig:1}
  \begin{center} %\hspace{0.3cm}
    \input{gaga.pstex_t}
     \caption{The diagrams contributing to the impact factor}
     \label{impact}
\end{center}
 \end{figure}
%%%%%%%%%%%%%%%%%%%%%%%%%%%%%%%%%%%%%%%%%%%%%%%%%%%%%%%%%%%%%%%%%%%%%%%%%%%%%
As usual, the integration of the two longitudinal components of the loop
momenta are absorbed into the definition of the impact factors, and the 
intermediate quark lines inside the impact factor are taken on-shell. 
Our task here is the calculation of the photon impact factor, and its 
interpretation in transverse coordinate space.

We first do the calculation in momentum space, and we compute the discontinuity
of the scattering amplitude. The assignment of the loop
momentum is illustrated in Fig.1; in all four graphs, the momenta 
of the cut quark lines are identical, e.g. $u-q$ for the upper quark line.    
We use Sudakov variable with the light cone vectors $p^\prime=p+yq$,
$y\simeq -p^2/2p\cdot q$  and $q^\prime=q+xp$,
$x\simeq -q^2/2p\cdot q$.
For the incoming and outgoing photon we introduce the invariants
\be
Q_{\pm}^2\,=\,-(q\pm l)^2,
~~~~~~~~~~s\equiv (p+q)^2\,\simeq\,2\,p^\prime \cdot q^\prime,
~~~~~~~~~~t\,=\,(2\,l)^2.
\ee
The other
photons involved in the process have similar expressions with the momentum $p$
instead of $q$. We write the momenta:
\begin{eqnarray}
  \label{eq:sud} \nonumber
  u\!\!&=&\!\! \alpha\, q^\prime\, +\, \beta\, p^\prime\, +\, u_\perp \\ \nonumber
  k\!\!&=&\!\!\alpha_k\, q^\prime\, +\, \beta_k\, p^\prime\, +\, k_\perp \\
  l\!\!&=&\!\! \alpha_l\, q^\prime\, + \,\beta_l\, p^\prime\, +\, l_\perp \,.
\end{eqnarray}
In terms of these parameters the photon virtualities are
\begin{equation}
  \label{eq:virt}
  {Q^2_\pm}\,\simeq\, (x\mp\beta_l) s- {l_\perp^2}.
\end{equation}
where we used the fact that
$\alpha_l\sim |t|/s\ll 1$ in the high energy limit.
In the following we use the Euclidian
form of the transverse momenta marked in boldface, i.e. $
k_\perp^2 = -{\bf k}^2<0$.
In the high energy limit, we also  approximate
the numerator of
the exchanged gluons  by the first term in the decomposition
\begin{equation}
  \label{eq:metr}
  g_{\mu\nu}\,=\,\frac2s (p_\mu^\prime q_\nu^\prime + p_\nu^\prime q_\mu^\prime ) +
  g_{\mu\nu}^\perp.
\end{equation}
With these simplifications, the integral of the first diagram takes the
following form:
\begin{eqnarray}
  \label{eq:int}
&&\hspace{-0.8cm}
\alpha_s \, {\rm e}^2 \sum_f q^2_f \,
  \frac{s}{2}\int \frac{d\beta_k}{2\pi} \int \frac{d\alpha\, d\beta \,
  d^2 u_\perp}{(2\pi)^4}
\\ \nonumber
\\ \nonumber
&&\hspace{-0.8cm}
\times\,
\frac{\delta [(\alpha -1)(\beta -x) s +
   u^2_\perp -m_f^2]\, \delta[\alpha (\beta - \beta_k)s+(u-k)^2_\perp -m_f^2]}
   {[\alpha (\beta -\beta_l)s+(u-l)^2_\perp -m_f^2][(\alpha -1)(\beta
  -\beta_k +x-\beta_l)s+(u-k-l)^2_\perp -m_f^2]}
\\ \nonumber
\\ \nonumber
&&\hspace{-0.8cm}
\times\,
 {\rm Tr}[\eslash (q+l)(\uslash 
 -\kslash -\qslash -\lslash +m_f^2)\pslash^\prime(\uslash-\qslash +m_f^2)
\eslash (q-l)(\uslash -\lslash +m_f^2)\pslash^\prime (\uslash -\kslash +m_f^2)] \, .
\end{eqnarray}
Performing the two $\beta$-integration with the delta functions leads to the following
relations
\begin{eqnarray}
  \beta_k\!\!&=&\!\!\frac{(u-k)^2_\perp -m^2_f}{\alpha
    s}+\frac{u^2_\perp -m_f^2}{(1-\alpha)s}-x \,
\\
\beta\!\!&=&\!\!\frac{u^2_\perp -m^2_f}{(1-\alpha)s}-x \, .
\end{eqnarray}
For longitudinally polarized photons, the polarization vectors read:
\begin{equation}
  \label{eq:photl}
  \epsilon_L(q\pm l)\,=\,\frac1{|Q_\pm|}\left[(1\pm\alpha_l)\, q^\prime
  +\left( x\mp\beta_l-\frac{2 l^2_\perp}{s}\right) p^\prime
\pm l_\perp\right]\,  ,
\end{equation}
The calculation of the trace in eq.~(\ref{eq:int}) greatly simplifies
if we make use of the Ward identity, i.e. the addition of a vector
proportional to $q\pm l$ to the polarisation vector does not
change the result. In this way the $q^\prime$ dependence of
$\epsilon_L$ can be eliminated.
The trace is computed in terms of Sudakov
variables using the fact that $p^\prime$
and $q^\prime$ are light cone vectors.
After summing the contributions corresponding to the four
diagrams in Fig.\ref{impact}, given by
the integrals of the type (\ref{eq:int}), the impact factor
for the longitudinally polarized photons takes the form:
\begin{eqnarray}
  \label{eq:ll}
\nonumber
\Phi_{LL}(q,k,l)
\!\!&=&\!\!
\alpha_s\sqrt{N_c^2-1} \; {\rm e}^2\, \sum_f q^2_f \,
4\sqrt{4\pi} \; |Q_+||Q_-|\,\int_0^1 d\alpha\int \frac{d^2{\bf u}}{(2\pi)^2}
\\ \nonumber
\\
\!\!&\times&\!\!
\alpha ^2 (1-\alpha )^2
\left( \frac1{D_1^+}-\frac1{D_2^+}\right)\left( \frac1{D_1^-}-
\frac1{D_2^-}\right)\, .
\end{eqnarray}
The four terms have been written in a factorised form, each of them
corresponds to one Feynman graph in Fig.\ref{impact}, where we have used:
\begin{eqnarray}
D_1^\pm\!\!&=&\!\!({\bf u}\pm (1-\alpha ) {\bf l})^2\,+\, \alpha (1-\alpha )\,Q_\pm^2\,+\,m_f^2
\\ \nonumber
\\
D_2^\pm\!\!&=&\!\!({\bf u-k}\,\mp\, \alpha{\bf l})^2\,\,+\,\alpha (1-\alpha )\,Q_\pm^2\,+\,m_f^2\,.
\end{eqnarray}

Setting $t=0$ one gets the well known expression in the forward direction.
In the same way we calculate the transverse impact factor where the
transverse polarisation vector reads:
\begin{equation}
  \label{eq:phott}
  \epsilon_T^{(h)} (q\pm l)\,=\,\epsilon^{(h)}_\perp
  \, \pm \,
  \frac{2
  l_\perp\cdot\epsilon^{(h)}_\perp}{s}\,\,(q^\prime-p^\prime\pm l_\perp)\,,
\end{equation}
where $h=\pm$ denotes two helicity states and
\begin{eqnarray} \label{eq:hel}
 \epsilon^{(h)}_\perp\, =\, \frac1{\sqrt 2}\,(0,1,\pm\, i,0)\,  .
\end{eqnarray}
%It is not possible to simplify the transverse polarisation vector with
%the Ward identity.
After some algebra and the use of the Ward identity,
the expression for the transverse impact factor becomes: 
\beeq
\label{eq:phitt}
\nonumber
&& \hspace{-1cm}
\Phi_{TT}^{(ij)}(q,k,l) \, =\, \alpha_s\, \sqrt{N_c^2-1}\; \mbox{\rm e}^2\,
\sum_f q^2_f \,
\sqrt{4\pi}\int_0^1 d\alpha \int \frac{d^2 {\bf u}}{(2\pi)^2}
\\ \nonumber
\\ \nonumber
&& \hspace{1.1cm}
\times\,
\left\{
-4\alpha(1-\alpha)\,\,\,
\epsbo_i \cdot
\left( \frac{{\bf N_1^+}}{D_1^+}-\frac{{\bf N_2^+}}{D_2^+}\right)
\left( \frac{{\bf N_1^-}}{D_1^-}-\frac{{\bf N_2^-}}{D_2^-}\right) \cdot
\epsbo_j^*
\right.
\\ \nonumber
\\
&& \hspace{-1cm}
\left.
+\,\,
\epsbo_i \cdot  \epsbo_j^*
\left[
\left( \frac{{\bf N_1^+}}{D_1^+}-\frac{{\bf N_2^+}}{D_2^+}\right)\cdot
\left( \frac{{\bf N_1^-}}{D_1^-}-\frac{{\bf N_2^-}}{D_2^-}\right)
+\,
m_f^2\,
\left( \frac1{D_1^+}-\frac1{D_2^+}\right)
\left( \frac1{D_1^-}-\frac1{D_2^-}\right)
\right]
\right\}.
\eeeq
where
\be
\epsbo_j=1/\sqrt{2} (1,\pm i)
\ee
are two-dimensional polarization vectors
corresponding to the two transverse polarizations $j=\pm$.
Again, it was possible to write the result in a factorised form where
we used the following definitions:
\begin{eqnarray}
{\bf N_1^\pm}\,=\,
{\bf u}\pm (1-\alpha){\bf l}\,~~~~~~~~~~~~~~~~~~
{\bf N_2^\pm}\!\!&=&\!\!{\bf u-k}\mp\alpha{\bf l}\,.
\end{eqnarray}
Setting $t=0$ and summing over the two helicity states,
one gets the well-known expressions in the forward direction, see
e.g. \cite{Soper}.  
This result also agrees with the one for real photons obtained in
\cite{Cheng-Wu}.

Finally, with the
same procedure one can study the impact factor for the incoming
photon longitudinally polarised while the outgoing
photon is transversely polarised.
\begin{eqnarray}
\label{eq:implt}
\nonumber
&& \hspace{-1cm}
\Phi_{LT}^{~(j)}(q,k,l)\,\,=\,\,\alpha_s\, \sqrt{N_c^2-1} \,\mbox{\rm  e}^2 \sum_f\,
q^2_f\, 2\, |Q_+| \sqrt{4\pi}
\int_0^1 d\alpha
\int \frac{d^2 {\bf u}}{(2\pi)^2}
\\ \nonumber
\\
&&\hspace{1.2cm}
\times\,\, \alpha
(1-\alpha )(1-2\alpha)
\left( \frac1{D_1^+}-\frac1{D_2^+}\right)\left( \frac{{\bf N_1^-}}{D_1^-}-
\frac{{\bf N_2^-}}{D_2^-}\right)\cdot{\epsbo}_j^*\,.
\end{eqnarray}
It is important to realize that  the impact factor $\Phi_{LT}$  is nonzero in the nonforward
direction in contrast to the forward case. This corrects the statement  made
in  the analysis   \cite{ForshawNonForw} in which $\Phi_{LT}=0$ in the nonforward case was claimed.
The reason for $\Phi_{LT}\ne 0$ is the symmetry property
of the integrand of  the integral over $\alpha$
in eq.~(\ref{eq:implt}). While in the forward case (${\bf l}=0$) the integrand
is antisymmetric with respect to the transformation $\alpha \rightarrow (1-\alpha)$,
thus giving zero after the integration over $\alpha$,
it becomes symmetric in the nonforward case (${\bf l}\ne 0$), leading to the nonzero result.

%%%%%%%%%%%%%%%%%%%%%%%%%%%%%%%%%%%%%%%%%%%%%%%%%%%%%%%%%%%%%%%%%%%%%%%%%%%%%%%%%%%%%
\section{Formulation in the coordinate space}
\label{sec:rspace}

It is instructive to transform these results to coordinate space.
We start from the general formulae:
\begin{eqnarray}
\label{eq:trans}
\nonumber
 \frac{1}{\kbo^2+\delta^2}
\!\!&=&\!\!
  \int \frac{d^2{\bf r}}{2\pi}\,e^{ i {\kbo \cdot \rbo}}\,
    K_0(\delta r)
\\ \nonumber
\\
\frac{\kbo}{\kbo^2+\delta^2}
\!\!&=&\!\!
\int \frac{d^2{\bf r}}{2\pi}\,e^{  i {\kbo \cdot \rbo}}\, i\,\delta   \,
\frac{\rbo}{r}\,
K_1(\delta r)\, .
\end{eqnarray}
where $K_{0,1} (z)$ are modified Bessel functions.
Let us concentrate first
on the longitudinal impact factor. After
using the first relation from the above, eq.~(\ref{eq:ll}) takes the form
\begin{eqnarray}
\nonumber
&&\hspace{-1cm}\Phi_{LL}(q,k,l)\,=\,\alpha_s\,\sqrt{N_c^2-1} \,
\mbox{\rm e}^2\, \sum_f q^2_f \,
4\sqrt{4\pi} \; |Q_+||Q_-|\, \int_0^1 d\alpha \, \alpha^2 (1-\alpha)^2 \,
\int \frac{d^2 \ubo}{(2\pi)^2}\,
\\  \nonumber
\\  \nonumber
&&\hspace{1cm} \times\; \int \frac{d^2{\bf r}}{2\pi}\;
e^{  i \ubo \cdot \rbo}
\left(e^{  i (1-\alpha)\lbo \cdot \rbo}-e^{ -i (\kbo+\alpha \lbo )\cdot \rbo}
\right) K_0(\delta_{+} r)\,
\\  \nonumber
\\
&&\hspace{1cm} \times\left( \;\int  \frac{d^2{\bf r^\prime}}{2\pi}\;
e^{  i \ubo \cdot \rbo^\prime}
\left(e^{ - i (1-\alpha)\lbo \cdot \rbo^\prime}-
e^{ -i (\kbo-\alpha \lbo )\cdot \rbo^\prime}
\right)\, K_0(\delta_{-} r^\prime) \right)^*
\end{eqnarray}
where
$
\delta_{\pm}^2=\alpha (1-\alpha) Q^2_{\pm}+m_f^2.
$
The integration over $\ubo$ leads to the delta function,
\beeq
\nonumber
\int \frac{d^2 \ubo}{(2\pi)^2}\; e^{  i \ubo \cdot (\rbo-\rbo^\prime)}
\,=\,\delta^2(\rbo-\rbo^\prime)\,,
\eeeq
which allows to perform the integration over $\rbo^\prime$. Thus we obtain
\begin{eqnarray}
\label{eq:ll1}
\nonumber
&&\hspace{-1.2cm}\Phi_{LL}(q,k,l)\,=\,\alpha_s\,\sqrt{N_c^2-1} \, \mbox{\rm e}^2\, \sum_f q^2_f \;
4\sqrt{4\pi} \; |Q_+||Q_-|\, \int_0^1 d\alpha \int \frac{d^{\/2} \rbo}{(2\pi)^2}
\\ \nonumber
\\
&&\hspace{-1cm} \times\;\alpha^2 (1-\alpha)^2
\underbrace{
e^{ i (1-\alpha) \lbo\cdot \rbo}\, K_0(\delta_{-} r)}
\,
\left( 1-e^{ -i({\kbo+\lbo)\cdot \rbo}}\right) \left(
1-e^{ i({\kbo-\lbo)\cdot \rbo}}\right) \,
\underbrace{
e^{ i (1-\alpha) \lbo\cdot \rbo}\,K_0(\delta_{+} r)}
\,.
\end{eqnarray}
It is important to note that the integrand (the second line) in (\ref{eq:ll1})
is invariant under the transformation:
$\alpha \to 1-\alpha$ and $\rbo \to - \rbo$. This reflects the symmetry of the
dipole formula under the interchange of quark and antiquark.

The underlined elements in (\ref{eq:ll1}) are proportional to the light-cone
wave functions of the longitudinal incoming and outgoing photons.
To be precise,
in the nonforward case the {\it longitudinal} photon wave function
$\Psi_{\lambda^\prime \lambda}^0$ (where $\lambda^\prime (\lambda )=\pm$
denotes the helicity of the (anti)quark) for a given flavour
$f$ is given by
\begin{eqnarray}
\label{eq:llwave}
 \nonumber
\Psi_{+-}^0(q\pm l,\rbo,\alpha)\,=\, \Psi_{-+}^0 \!\!\!&=&\!\!\!
\frac{e\, q_f}{2\,\pi^{3/2}}\, \sqrt{N_c}\,\alpha (1-\alpha)\,
|Q_{\pm}|\, K_0(\delta_{\pm} r) \,
e^{ \pm i (1-\alpha) \lbo\cdot \rbo}\,
\\ \nonumber
\\
\Psi_{++}^0\,=\,\Psi_{--}^0\!\!\! &=&\!\!\! 0 \,.
\end{eqnarray}
The above wave function differs from the photon wave function with only 
longitudinal momentum,
known from the total cross section for the $\gamma^*\gamma^*$
scattering~\cite{Nikolaev},
by the exponential factor involving transverse momentum $\lbo$ and the 
longitudinal momentum fraction $\alpha$ (this factor has also been found in 
~\cite{Wusthoff-Ivanov,IK}). It has important consequences for the discussion
of the  impact parameter representation presented in the next section.
Summarizing, the impact factor (\ref{eq:ll1}) in the coordinate space becomes
\beeq
\label{eq:ll2}
\nonumber
&&\hspace{-2.0cm}\Phi_{LL}(q,k,l)\,=\, 4 \pi^{3/2}\, 
\alpha_s\,\frac{\sqrt{N_c^2-1}}{N_c}\,
\int_0^1 d\alpha\, \int {d^{\/2} \rbo}\,\,
\left(1-e^{ -i({\kbo+\lbo)\cdot \rbo}}\right)
\left(
1-e^{ i({\kbo-\lbo)\cdot \rbo}}\right) \,
\\  \nonumber
\\
&&\hspace{0cm}  \times\;
\sum_f\, \sum_{\lambda^\prime \lambda}\,\left\{
\overline{\Psi}_{\lambda^\prime \lambda}^0(q-l,\rbo,\alpha)\,
\Psi_{\lambda^\prime \lambda}^0(q+l,\rbo,\alpha)
\right\}\,.
\eeeq
where $\overline{\Psi}$ is  complex conjugate to ${\Psi}$.
Inserting this relation into the formula (\ref{eq:a})
for the $\gamma^*\gamma^*$ scattering amplitude, 
we find for the longitudinally polarized external photons
\beeq
\label{eq:adip}
\nonumber
&&\hspace{-2.5cm}
A_{LL}(s,\lbo)\,=\,
is
\int {d^{\/2} \rbo_1}   \int {d^{\/2} \rbo_1}
\int_0^1 d{\alpha}_1  \int_0^1 d{\alpha}_2
\\   \nonumber
\\   \nonumber
&&\hspace{-0.8cm} \times\,
%\left\{
\sum_{f_1}\, \sum_{\lambda^\prime \lambda}\, \left\{
\overline{\Psi}_{\lambda^\prime \lambda}^0(q-l,\rbo_1,\alpha_1)\,
\Psi_{\lambda^\prime \lambda}^0(q+l,\rbo_1,\alpha_1)
\right\}
\\   \nonumber
\\   \nonumber
&&\hspace{-0.8cm} \times\,\,{N}(\rbo_1,\rbo_2,\lbo)
\\   \nonumber
\\
&&\hspace{-0.8cm} \times
%\left\{
\sum_{f_2}\,\sum_{\lambda^\prime \lambda}\,
\left\{
\overline{\Psi}_{\lambda^\prime \lambda}^0(p+l,\rbo_2,\alpha_2)\,
\Psi_{\lambda^\prime \lambda}^0(p-l,\rbo_2,\alpha_2)
\right\}\,,
\eeeq
where $N(\rbo_1,\rbo_2,\lbo)$ is the scattering amplitude  of two dipoles
of the transverse size
$\rbo_1$ and $\rbo_2$  with the momentum transfer $2\lbo$,
in the two gluon exchange approximation
\beeq
\label{eq:dd}
\nonumber
&&\hspace{-2.0cm}
N(\rbo_1,\rbo_2,\lbo)\,= \alpha_s^2 \,\frac{(N_c^2-1)}{N_c^2}
\int \frac{d^2\kbo}{(\kbo+\lbo)^2(\kbo-\lbo)^2}
\\ \nonumber
\\
&&\hspace{0.0cm} \times\,
\left( 1-e^{ -i({\kbo+\lbo)\cdot \rbo_1}}\right)
\left(1-e^{ i({\kbo-\lbo)\cdot \rbo_1}}\right)
\left( 1-e^{ i({\kbo+\lbo)\cdot \rbo_2}}\right)
\left(1-e^{ -i({\kbo-\lbo)\cdot \rbo_2}}\right)\,.
\eeeq

Let us concentrate now on the impact factor for {\it transverse} photons
$\Phi_{TT}^{(ij)}$,
eq.~(\ref{eq:phitt}).
Using formulas (\ref{eq:trans}) and repeating the steps presented in the
longitudinal case, we obtain
\beeq
\label{eq:phittnew}
\nonumber
&&\hspace{-1.0cm}
\Phi_{TT}^{(ij)}(q,k,l)\, = \,\alpha_s\, \sqrt{N_c^2-1}\,e^2
\sum_f q_f^2\,\,
\sqrt{4\pi}\int_0^1 d\alpha \int {\frac{d^2{\bf r}}{(2\pi)^2}}\,\,
\left( 1-e^{ -i({\kbo+\lbo)\cdot \rbo}}\right) \left(
1-e^{ i({\kbo-\lbo)\cdot \rbo}}\right)
\\ \nonumber
\\ \nonumber
&&\hspace{1.1cm} \times\,
\biggl\{
-4\alpha(1-\alpha)
\left(
\frac{\epsbo_j \cdot {\bf r}}{r}
\delta_{-} K_1(\delta_{-} r)\, e^{-i (1-\alpha) \lbo\cdot \rbo}\right)^*
\left(
\frac{\epsbo_i \cdot {\bf r}}{r}
\delta_{+} K_1(\delta_{+} r)\,e^{i (1-\alpha) \lbo\cdot \rbo}\right)
%\right.
\\ \nonumber
\\ \nonumber
&&\hspace{1.1cm} +\,\,
%\left.
\delta_{ij}\,
\left(\delta_{-}\,K_1(\delta_{-} r)\, e^{-i (1-\alpha) \lbo\cdot \rbo}\right)^*
\left(\delta_{+}\,K_1(\delta_{+} r)\, e^{i (1-\alpha) \lbo\cdot \rbo}\right)
%\right.
\\ \nonumber
\\
&&\hspace{1.1cm} +\,\,
%\left.
\delta_{ij}\, m_f^2
\left(K_0(\delta_{-} r)\, e^{-i (1-\alpha) \lbo\cdot \rbo}\right)^*
\left(K_0(\delta_{+} r)\, e^{i (1-\alpha) \lbo\cdot \rbo}\right)
\biggr\}
\eeeq
where we used the fact that $\epsbo_i \cdot  \epsbo_j^*=\delta_{ij}$.
The above formula can be written in a similar form as eq.~(\ref{eq:ll2})
using helicity wave functions
\beeq
\label{eq:phittnew1}
\nonumber
\hspace{-0cm}
\Phi_{TT}^{(ij)}(q,k,l) &=& 4 \, \pi^{3/2}\, 
\alpha_s\,\frac{\sqrt{N_c^2-1}}{N_c}\,\int_0^1 d\alpha \int {d^2 {\bf r}}\,\,
\left( 1-e^{-i({\kbo+\lbo)\cdot \rbo}}\right) \left(
1-e^{i({\kbo-\lbo)\cdot \rbo}}\right)
\\ \nonumber
\\
&\times& \sum_f \, \sum_{\lambda^\prime \lambda}
\left\{
\overline{\Psi^j_{\lambda^\prime \lambda}}(q-l,\rbo,\alpha)\,
\Psi^i_{\lambda^\prime \lambda}(q+l,\rbo,\alpha)
\right\}
\eeeq
where $i,j$ are the helicities of the incoming and outgoing photon,
respectively, and  $\lambda^\prime (\lambda )$
denotes the helicity of the (anti)quark.
The photon wave functions are defined in terms of these helicities as:
\beeq
\label{eq:wftt}
\Psi_{+-}^{+}(q\pm l,\rbo,\alpha)
\!\!&=&\!\! \frac{i\,e\, q_f}{2\,\pi^{3/2}}\,\sqrt{N_c}\, \alpha\,
\frac{\epsbo_+\cdot\rbo}{r}\,\delta_{\pm}\,K_1(\delta_{\pm} r)
\,e^{\pm i (1-\alpha) \lbo\cdot \rbo}
\\   \nonumber
\\
\label{eq:wftt29}
\Psi_{+-}^{-}(q\pm l,\rbo,\alpha)
\!\!&=&\!\!-\frac{i\,e\, q_f}{2\,\pi^{3/2}}\,\sqrt{N_c}\, (1-\alpha)\,
\frac{\epsbo_-\cdot\rbo}{r}\,\delta_{\pm}\,K_1(\delta_{\pm} r)
\,e^{\pm i (1-\alpha) \lbo\cdot \rbo}
\\   \nonumber
\\
\label{eq:wftt30}
\Psi_{-+}^{+}(q\pm l,\rbo,\alpha)
\!\!&=&\!\! -\frac{i\,e\, q_f}{2\,\pi^{3/2}}\,\sqrt{N_c}\, (1-\alpha)\,
\frac{\epsbo_+\cdot\rbo}{r}\,\delta_{\pm}\,K_1(\delta_{\pm} r)
\,e^{\pm i (1-\alpha) \lbo\cdot \rbo}
\\   \nonumber
\\
\label{eq:wftt31}
\Psi_{-+}^{-}(q\pm l,\rbo,\alpha)
\!\!&=&\!\!\frac{i\,e\, q_f}{2\,\pi^{3/2}}\,\sqrt{N_c}\, \alpha\,
\frac{\epsbo_-\cdot\rbo}{r}\,\delta_{\pm}\,K_1(\delta_{\pm} r)
\,e^{\pm i (1-\alpha) \lbo\cdot \rbo}
\\  \nonumber
\\
\label{eq:wftt32}
\Psi_{++}^{+}=\Psi_{--}^{-}(q\pm l,\rbo,\alpha)
\!\!&=&\!\!  \frac{e\, q_f}{(2\pi)^{3/2}}\, \sqrt{N_c}\,
m_f\,K_0(\delta_{\pm} r)\,e^{\pm i (1-\alpha) \lbo\cdot \rbo}
\\   \nonumber
\\
\Psi_{--}^{+}=\Psi_{++}^{-}\!\!&=&\!\! 0 \, .
\eeeq
Again, the exponentials in the above are due to the transverse momentum
of the photons. Similar wave functions were defined in
\cite{Wusthoff-Ivanov} and  \cite{Gieseke-Qiao}. The difference
between reference ~\cite{Wusthoff-Ivanov} and
our results lies in the minus sign in (\ref{eq:wftt29}) and
(\ref{eq:wftt30}), and we found a different 
relative normalization between  (\ref{eq:wftt32})  and the rest of the nonzero
wave functions (\ref{eq:wftt})--(\ref{eq:wftt31}) compared to \cite{Gieseke-Qiao}.
It is important to realize that in the forward case, when $\lbo=0$ and $i=j$,
after the summation over  the two transverse polarizations,
the expression in the curly brackets in eq.~(\ref{eq:phittnew}) becomes
a well known expression for the square of the photon wave function in
forward kinematics
\beeq
\nonumber
|\Psi_T(q,\rbo,\alpha)|^2\,\sim\,
\sum_f e_f^2 \left\{[\alpha^2+(1-\alpha)^2 ]\,\delta^2\,K_1^2(\delta r)\,+\,
m_f^2\,K_0^2(\delta r)
\right\}
\eeeq
where $\delta^2=\alpha (1-\alpha) Q^2+m_f^2$.   The structure of   the full
$\gamma^*\gamma^*$ amplitude for transverse photons
is the same as in eq.~(\ref{eq:adip}) with the
wave function factors replaced by those in the curly brackets in eq.~(\ref{eq:phittnew1}).

For completeness we also  present the formula for the impact factor with mixed
polarization (\ref{eq:implt})
\begin{eqnarray}
\label{eq:impltnew}
\Phi_{LT}^{~(j)}(k,l) &=& 4\, \pi^{3/2}\, 
\alpha_s\,\frac{\sqrt{N_c^2-1}}{N_c}\,
\int_0^1 d\alpha
\int {d^2 {\bf r}}\,
\left( 1-e^{ -i({\kbo+\lbo)\cdot \rbo}}\right) \left(
1-e^{ i({\kbo-\lbo)\cdot \rbo}}\right)
\\ \nonumber
\\ \nonumber
&\times& \sum_f \, \sum_{\lambda^\prime \lambda}
\left\{
\overline{\Psi^j_{\lambda^\prime \lambda}}(q-l,\rbo,\alpha)
\Psi^0_{\lambda^\prime \lambda}(q+l,\rbo,\alpha)\,
\right\}
\end{eqnarray}
In each presented case  the $\gamma^*\gamma^*$ scattering amplitude $A(s,\lbo)$ has the form
of eq.~(\ref{eq:adip})  with the wave function replacement  which takes into account
helicities of the external photons.

%%%%%%%%%%%%%%%%%%%%%%%%%%%%%%%%%%%%%%%%%%%%%%%%%%%%%%%%%%%%%%%%%%%%%%%%%%%%%%%%%%%%%%%%%%%%
\section{Impact parameter}

In our notation,
the four-momentum transfer squared $t$ between the two  $q\bar{q}$ systems is given
by the transverse two-dimensional vector $\Deltabo=2\lbo$, i.e. $t=-\Deltabo^2$.
The impact parameter $\bbo$ is defined as a Fourier conjugate
variable to $\Deltabo$, and the $\gamma^*\gamma^*$
scattering amplitude in the impact parameter space is given by
\be
\label{eq:ampb}
\tilde{A}(s,\bbo)\,=\,\int \frac{d^{\/2}\Deltabo}{(2\pi)^2}\,
{e}^{\textstyle i\bbo\cdot\Deltabo}\,
A(s,\Deltabo)\,.
\ee
Performing this integral, we keep the total energy $\sqrt{s}$  as well as
virtualities of the external photons, $Q^2_\pm~ \mbox{\rm and}~ P^2_\pm$, fixed.
Let us concentrate on the amplitude for longitudinal photons (\ref{eq:adip})
with the wave function (\ref{eq:llwave}). The dependence on  $\Deltabo$ (or $\lbo$)
resides in  the exponential factors in the wave functions and  the dipole-dipole
scattering amplitude  $N(\rbo_1,\rbo_2,\Deltabo)$,
eq.~(\ref{eq:dd}). Thus performing the transformation  (\ref{eq:ampb}) the
amplitude (\ref{eq:adip}) we find
\beeq
\label{eq:dint}
\nonumber
&&\hspace{-2.5cm}
\tilde{A}(s,\bbo)\,=\,
is
\int {d^{\/2} \rbo_1}   \int {d^{\/2} \rbo_1}
\int_0^1 d{\alpha}_1  \int_0^1 d{\alpha}_2
\\   \nonumber
\\   \nonumber
%\left\{
&&\hspace{-1.0cm} \times\, \sum_{f_1}\,\sum_{\lambda^\prime \lambda}
\left\{\overline{\Psi}_{\lambda^\prime\lambda}(Q_{-},\rbo_1,{\alpha}_1)\,
\Psi_{\lambda^\prime\lambda}(Q_{+},\rbo_1,{\alpha}_1)
\right\}
\\   \nonumber
\\   \nonumber
&&\hspace{-1.0cm} \times\,\, \tilde{N}(\rbo_1,\rbo_2,\bbo+(1-\alpha_1) 
\rbo_1+(1-\alpha_2) \rbo_2)
\\   \nonumber
\\
&&\hspace{-1.0cm} \times\,
%\left\{
\sum_{f_2}\,\sum_{\lambda^\prime \lambda}
\left\{ \overline{\Psi}_{\lambda^\prime\lambda}(P_{+},\rbo_2,{\alpha}_2)\,
\Psi_{\lambda^\prime\lambda}(P_{-},\rbo_2,{\alpha}_2)
\right\}\,.
\eeeq
The wave functions above are the forward photon wave functions with the indicated
photon virtualities, given by eqs.~(\ref{eq:wftt}) with the exponential 
factors being removed.
This is due to the fact that, when transforming to impact parameter,
we have to include the momentum transfer 
dependence of these factors into the  definition of the 
dipole-dipole scattering amplitude $\tilde{N}$: 
\beeq
\label{eq:ddimp}
\nonumber
&&\hspace{-1.5cm}
\tilde{N}(\rbo_1,\rbo_2,\bbo+(1-\alpha_1)\rbo_1 + (1-\alpha_2) \rbo_2)
\\ \nonumber
\\
&&\hspace{0cm}
= \int \frac{d^{\/2}\Deltabo}{(2\pi)^2}\,\,
{e}^{\textstyle i
\left(\bbo+ (1-\alpha_1)\rb+ (1-\alpha_2)\rbb\right)\cdot\Deltabo}\,\,
N(\rbo_1,\rbo_2,\Deltabo)\,.
\eeeq
From (\ref{eq:ddimp}) we see that, in the dipole-dipole  scattering amplitude,
the dependence upon the impact parameter contains an $\alpha$-dependent shift 
relative to the impact parameter $\bbo$ which refers to the external photon
eq.~(\ref{eq:ampb}): $\bbo\to \bbo+ (1-\alpha_1)\rbo_1+ (1-\alpha_2)\rbo_2$.
Furthermore, as we have already remarked after eq.~(\ref{eq:ll1}), the dipole
scattering amplitude (for fixed transverse dipole sizes and longitudinal momenta)
is invariant under the interchanges
$\alpha_1 \to (1-\alpha_1)$,
$\rbo_1 \to - \rbo_1$ or/and $\alpha_2 \to (1-\alpha_2)$,
$\rbo_2 \to - \rbo_2$.

%%%%%%%%%%%%%%%%%%%%%%%%%%%
\begin{figure}[t]
  \vspace*{-1cm}
     \centerline{
         \epsfig{figure=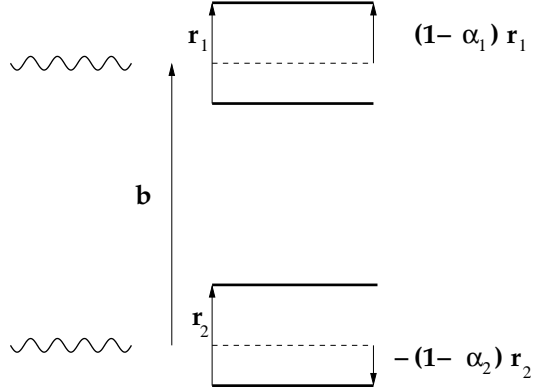,width=7cm}
           }
\vspace*{0.cm}
\caption{\it  The interpretation of relation (\ref{eq:bvec}) in the
transverse coordinate plane: 
the wavy lines denote the incoming photons, the full lines the quark
antiquarks of the upper and the lower color dipoles.}
\label{fig:dipol2}
\end{figure}
%%%%%%%%%%%%%%%%%%%%%%%%%%%%%%

Let us give a physical interpretation of this result.
In the high energy limit, the incoming photon and the quark-antiquark pair  
are conveniently described 
in an infinite momentum frame. In this frame, as it has been shown many years 
ago \cite{kogut,Soper2}, there exists a subgroup of the Poincare group which is
isomorphic to the symmetry group of Galilei transformations in 
nonrelativistic two-dimensional quantum mechanics. This two dimensional 
motion takes place in the transverse plane, and the longitudinal momentum 
plays the role of the nonrelativistic mass. In our case of dipole-dipole 
scattering we choose the upper incoming photon to move in the positive
z-direction, i.e. in light cone variables $q$ has a large $q_+$ component.
Inside the upper impact factor, the upper quark line with longitudinal 
momentum
fraction $1-\alpha_1$ carries the 'mass' $m_u=(1-\alpha_1)q_+$, whereas the 
lower
quark line has the 'mass' $m_l=\alpha_1 q_+$. The vector $\rbo_1$ denotes the
transverse distance between the upper and lower quark line, and the impact 
parameter $\bbo$ is the transverse distance between the two 
incoming photons. If we interpret the upper incoming photon as being in the 
center of mass of the upper quark-antiquark system, see Fig.~\ref{fig:dipol2}, 
the two vectors
\beeq
&&\hspace{-1cm}
\bbo+\frac{m_u}{m_u+m_l} \rbo_1\, =\, \bbo+(1-\alpha_1)\rbo_1
\eeeq
and
\beeq
&&\hspace{-2cm}
\bbo-\frac{m_l}{m_u+m_l}\rbo_1\, =\, \bbo-\alpha_1 \rbo_1
\eeeq
denote the position of the upper and lower quark, respectively.  A similar argument
holds for the lower dipole, i.e. the vector 
\be
\label{eq:bvec}
\bbo + (1-\alpha_1)\, \rbo_1 + (1-\alpha_2)\, \rbo_2 
\ee
denotes the distance between the upper quark line in the upper dipole and
the lower quark line in the lower dipole.  
The peculiar $\bbo$-dependence of eq.~(\ref{eq:ddimp}) then says that the
interaction between the upper and lower color dipoles depends upon 
the distance between one of the two quarks of the upper 
dipole and one of the quarks in the lower dipole (and not upon the distance 
$\bbo$ between the center of mass coordinates of the two dipoles).
Moreover, because of the symmetry under 
$\alpha \to 1-\alpha$, $\rbo \to -\rbo$ we are free to interchange the 
quarks inside one of the two color dipoles.

Finally, the appearance of the phase factors $\exp\{\pm i(1-\alpha)\lbo\cdot \rbo\}$
in (\ref{eq:wftt}) can be understood as follows. Inverting (19) we 
find, for example, for the upper incoming longitudinal photon    
\beeq
K_0(\delta_+ r)\, e^{i(1-\alpha) \lbo \rbo}\, =\,
        \int \, \frac{d^2 \kbo}{2 \pi} \, e^{-i(\kbo - (1-\alpha)\lbo)\cdot \rbo}\,
           \frac{1}{\kbo ^2 +\delta_+^2}\,,
\eeeq
i.e. the momentum conjugate to the transverse distance $\rbo$ is
$(\kbo - (1-\alpha)\lbo)$ rather than $\kbo$. The incoming photon carries
transverse momentum $\lbo$ and splits into quark and antiquark with 
longitudinal and transverse momenta $(1-\alpha$, $\kbo)$ and
$(\alpha$, $-\kbo+\lbo)$, respectively. In the infinite momentum frame this
corresponds to the masses $m_1=(1-\alpha)q_+$ and $m_2=\alpha q_+$.
From nonrelativistic mechanics
we know that the relative momentum of the quark-antiquark pair
(i.e. the momentum conjugate to the relative coordinate $\rbo$) is given by
\be
\frac{m_2}{m_1 +m_2} \pbo_1 - \frac{m_1}{m_1 +m_2} \pbo_2\,=\,
\alpha \kbo -(1-\alpha)(-\kbo +\lbo)= \kbo -(1-\alpha) \lbo\,.
\ee
So the appearance of the phase factor looks very natural.

%%%%%%%%%%%%%%%%%%%%%%%%%%%%%%%%%%%%%%%%%%%%%%%%%%%%%%%%%%%%%%%%%%%%%%%%%%%%%%%%
\section{Nonforward dipole-proton scattering amplitude}

In a case when the lower photon is replaced by the proton, an unintegrated
gluon distribution $f(x_\pm,,\kbo,\lbo)$ appears \cite{Jeffbook}, where
$x_\pm=Q^2_{\pm}/s$ and the adopted notation means that
$f$ depends on both the variables. Thus, strictly speaking, $f$ is a nondiagonal (skewed or generalized)
unintegrated gluon distribution since both the longitudinal and transverse
momenta of the exchanged gluons, $k\pm l$, are not equal,
see Fig.~\ref{fig:1}.
In the proton case,  the amplitude (\ref{eq:a}) reads
\be
\label{eq:anew}
A(s,\lbo=\Deltabo/2)\,=\,\frac{is}2 \int
\frac{d^2{\bf k }}{(2\pi)^3}\,\frac{\Phi(q,\kbo,\lbo)}{{\bf (k+l)}^2}
\frac{f(x_\pm,\kbo,\lbo)}{{\bf (k-l)}^2 } \, .
\ee
Repeating the steps from  section \ref{sec:rspace}, we find in the coordinate space
the analogue of eq.~(\ref{eq:adip})
\be
\label{eq:adipnew}
A(s,\lbo=\Deltabo/2)=
is
\int {d^{\/2} \rbo}
\int_0^1 d{\alpha}\,
\sum_{f} \,
\sum_{\lambda^\prime \lambda}
\left\{\overline{\Psi}_{\lambda^\prime\lambda}(q-l,\rbo,{\alpha})\,
{N}(x_\pm,\rbo,\lbo)\,
\Psi_{\lambda^\prime\lambda}(q+l,\rbo,{\alpha})
\right\}
\ee
where now ${N}(x_\pm,\rbo,\lbo)$  is a nonforward dipole--proton
scattering amplitude
\beeq
\label{eq:adiprot}\nonumber
{N}(x_\pm,\rbo,\lbo) &=& \frac{\alpha_s}{4\pi^{3/2}}\,  \frac{\sqrt{N_c^2-1}}{N_c} 
\\
&\times& 
\int \frac{d^2\kbo}{(\kbo+\lbo)^2(\kbo-\lbo)^2}
\left( 1-e^{ -i({\kbo+\lbo)\cdot \rbo}}\right)
\left(1-e^{ i({\kbo-\lbo)\cdot \rbo}}\right) f(x_\pm,\kbo,\lbo)\,.
\eeeq
In the impact parameter representation we have:
\be
\label{eq:adipnew1}
\tilde{A}(s,\bbo)\,=\,
is\int {d^{\/2} \rbo}
\int_0^1 d{\alpha}\,
\sum_{f} \,
\sum_{\lambda^\prime \lambda}
\left\{\overline{\Psi}_{\lambda^\prime\lambda}(Q_-,\rbo,{\alpha})\,
\tilde{N}(x_\pm,\rbo,\bbo+ (1-\alpha)\rbo)\,
\Psi_{\lambda^\prime\lambda}(Q_+,\rbo,{\alpha})
\right\}
\ee
with
\be
\label{eq:nb}
\tilde{N}(x_\pm,\rbo,\bbo+ (1-\alpha)\rbo)      1
\,=\,
\int \frac{d^{\/2}\Deltabo}{(2\pi)^2}\,
{e}^{\textstyle i
\left(\bbo+ (1-\alpha)\rbo \right)\cdot\Deltabo}\,
{N}(x_\pm,\rbo,\lbo=\Deltabo/2)\,.
\ee
Again, there are forward photon wave functions in (\ref{eq:adipnew1})  since
the nonforward exponentials are incorporated into   (\ref{eq:nb}).
As we have mentioned before, this expression is invariant under the
replacement: $\alpha \to 1-\alpha$ and  $\rbo \to -\rbo$, which can be easily seen
by inserting  (\ref{eq:adiprot}) into (\ref{eq:nb}) and performing the symmetry transformation.

For completeness we also quote the formula for diffractive vector 
production in the nonforward direction ~\cite{IK}, e.g. for the 
production of longitudinal $\rho$-mesons. In our kinematics, the impact 
factor has the form 
(\ref{eq:adipnew}), with the wave function for the outgoing photon,
$\Psi(q-l,\rbo,\alpha)$, being replaced by the meson wave function:
\be
\Psi(q-l,\rbo,\alpha)\,\, \,\to\,\,\, \Psi_{\rho}(\rbo,\alpha)\, e^{-i(1-\alpha)\lbo \rbo}
\ee
As in the case of $\gamma^*\gamma^*$ scattering, the extra phase factor 
accounts for the nonzero transverse momentum of the vector particle.        
%%%%%%%%%%%%%%%%%%%%%%%%%%%%%%%%%%%%%%%%%%%%%%%%%%%%%%%%%%%%%%%%%%%%%%%%%%%%
\section{Open quark-antiquark production}

In the final step of our discussion we remove, in the upper
dipole system, the wave function of the outgoing photon, i.e. we consider 
the diffractive production of an open quark-antiquark pair in the nonforward 
direction, $\gamma^* + p\to (q\bar{q})+p$.
We follow our previous
notation: the incoming photon carries the momentum $q+l$, and the
outgoing antiquark and quark have the momenta $q-u$ and $u-l$, respectively.
The scattering amplitude has the form (\ref{eq:anew}) with the replacement:
$\Phi(q,k,l) \to \Phi_{\lambda^\prime\lambda}^{\gamma^* \to (q\bar{q})} (q,u,k,l)$. For the
longitudinal photon, the new impact factor for open quark-antiquark production
is given by:
\beeq
\label{eq:qq}
\nonumber
&&\hspace{-2cm}
\Phi^{0,\gamma^* \to (q\bar{q})}_{+-}(q,\ubo,\kbo,\lbo) \,=\, 
2\,(2\pi)^{3/2} \alpha_s \sqrt{\frac{N_c^2-1}{N_c}}\, e_f\, [\alpha (1-\alpha)]^{3/2}\, |Q_+|
\\ \nonumber
\\
&&\hspace{-1.5cm}
\times
\left( \frac{1}{D(\ubo+(1-\alpha)\lbo)}
            +\frac{1}{D(\ubo-(1+\alpha)\lbo)}
            -\frac{1}{D(\ubo+\kbo-\alpha\lbo)}
            -\frac{1}{D(\ubo-\kbo-\alpha\lbo)} \right)
\eeeq
with $D(\kbo)= \kbo^2 + \alpha (1-\alpha)Q_+^2 = \kbo^2 + \delta_+^2$.
Inserting (\ref{eq:trans}), we arrive at:
\beeq
\nonumber
&&\hspace{-2cm}
\Phi^{0,\gamma^* \to (q\bar{q})}_{+-}(q,\ubo,\kbo,\lbo) \,=\, 
2\,(2\pi)^{3/2} \alpha_s \sqrt{\frac{N_c^2-1}{N_c}}\, e_f\, [\alpha (1-\alpha)]^{3/2}\, |Q_+|
\\ \nonumber
\\
&&\hspace{0.0cm}
\times
\int \frac{d^2 \rbo'}{2 \pi}\,e^{i\ubo\cdot\rbo'}
         \left( 1-e^{-i(\kbo + \lbo)\cdot \rbo'} \right)
         \left( 1-e^{i(\kbo - \lbo)\cdot \rbo'} \right)
          e^{i(1-\alpha)\lbo \cdot \rbo'} K_0(\delta_+r')\,.
\eeeq

From the discussion at the end of section 3 we know that the
transverse distance between the outgoing quark an antiquark, $\rbo$,
is conjugate to the transverse momentum $\ubo -(1-\alpha)\lbo$.
We therefore define the scattering amplitude for the production of a dipole
of the size $\rbo$ by taking the Fourier transform
\be
\Phi^{0,\gamma^* \to (q\bar{q})}_{+-}(q,\rbo,\kbo,\lbo)\,\equiv\,
\int \frac{d^2 \ubo}{(2\pi)^2}\, e^{-i(\ubo - (1-\alpha)\lbo)\cdot \rbo}\,
\Phi^{0,\gamma^* \to (q\bar{q})}_{+-}(q,\ubo,\kbo,\lbo)\,,
\ee
and
\beeq
\nonumber
\hspace{-0cm}\Phi^{0,\gamma^* \to (q\bar{q})}_{+-}(q,\rbo,\kbo,\lbo)
\!\!\!&=&\!\!\!
%&&\hspace{-0.6cm}
2\,\sqrt{2\pi}\, \alpha_s \sqrt{\frac{N_c^2-1}{N_c}}\, e_f\,
[\alpha (1-\alpha)]^{3/2}\, |Q_+|\,
\\ \nonumber
\\ \nonumber
%&&\hspace{0cm}
&\times&\!\!\!
e^{i(1-\alpha)\lbo \cdot \rbo}
\left( 1-e^{-i(\kbo + \lbo)\cdot \rbo} \right)
         \left( 1-e^{i(\kbo - \lbo)\cdot \rbo} \right)
e^{i(1-\alpha)\lbo \cdot \rbo}\, K_0(\delta_+r)
\\ \nonumber
\\ \nonumber
%&& \hspace{-0.6cm}
&=&\!\!\!
\sqrt{2}\,(2\pi)^2 \alpha_s \frac{\sqrt{N_c^2-1}}{N_c}
\sqrt{\alpha (1-\alpha)}
\\ \nonumber
\\
%&&\hspace{-0.6cm}
&\times&\!\!\!
e^{i(1-\alpha)\lbo \cdot \rbo}
\left( 1-e^{-i(\kbo + \lbo)\cdot \rbo} \right)
         \left( 1-e^{i(\kbo - \lbo)\cdot \rbo} \right) \Psi_{+-}^0 (q+l,\rbo,\alpha).
\eeeq
Finally, the scattering amplitude of the dipole on a proton our
result takes the form:
\beeq
A_{+-}^{\gamma^* \to (q\bar{q})}(s,\alpha, \rbo,\lbo) \,=\, is\,
\sqrt{2\pi}\, \sqrt{\alpha(1-\alpha)}\, e^{i(1-\alpha)\lbo \cdot \rbo}\,
N(x_\pm,\rbo,\lbo)\, \Psi_{+-}^0(q+l,\rbo,\alpha).
\eeeq
with the dipole scattering amplitude from (\ref{eq:adiprot}).
When transforming to impact parameter $\bbo$, again the shifted variable
$\bbo +(1-\alpha)\lbo$ appears, the distance from the target center
to the quark. For the integrated and spin summed cross section
\beeq
\frac{d\sigma}{d^2 \Deltabo}\, =\,\frac{1}{16 \pi^2 s^2}\,\frac{1}{4\pi}\,
\frac1{\alpha(1-\alpha)}\int_0^1 d\alpha \int {d^2\ubo\over(2\pi)^2}\,
\sum_{\lambda^\prime\lambda}|A_{\lambda^\prime\lambda}^{\gamma^* \to (q\bar{q})}(s,\alpha,\ubo,\lbo=\Deltabo/2)|^2\,,
\eeeq
we obtain after transforming to transverse coordinates:
\beeq
\frac{d\sigma}{d\Deltabo^2}\, =\,\frac{1}{16 \pi} \int_0^1 d \alpha \int d^2 \rbo\,\,\sum_{\lambda^\prime\lambda}
        \overline{{\Psi}_{\lambda^\prime\lambda}^0}(q+l,\rbo,\alpha)\, |N(x_\pm,\rbo,\lbo)|^2\,
             \Psi_{\lambda^\prime\lambda}^0(q+l,\rbo,\alpha).
\eeeq
This generalizes the well-known formula for the diffractive 
quark-antiquark production in the forward direction ~\cite{Nikolaev}.
In particular, the integrated cross section for diffractive quark-antiquark 
production contains the same dipole scattering amplitude as 
the elastic $\gamma^* p$ process (45).   
%When 
%transforming to imapct parameter $\bbo$, it is convenient to write
%\be
%|N(x_\pm,\rbo,\lbo)|^2 = N(x_\pm,\rbo,\lbo)^* e^{-i(1-\alpha)\rbo \lbo}
%  \;\;\cdot  e^{i(1-\alpha)\rbo \lbo} N(x_\pm,\rbo,\lbo)
%\ee
\section{Conclusions}

In this paper we have discussed the generalization of the color dipole picture 
to the nonforward direction. For the elastic scattering of two color dipoles  
we found that the dipole cross section depends upon the transverse distance
between the ends of the two dipoles, and not between the centers of the 
two quark-antiquark pairs. 
This result also holds for diffractive vector production. For open 
$q\bar{q}$-production, we confirm the well-known result of the forward
direction: the integrated (over the transverse size of the produced
quark-antiquark pair) cross section is described by the square of the
elastic dipole cross section. 

We believe that our findings might be useful in several respects.
First, it may provide some guidance in  
modelling the $b$-dependence of the dipole cross section. As we have
indicated in our introduction, the $b$-dependence represents the mayor
challenge in understanding the transition from the region of perturbative 
to nonperturbative QCD, and our formula may provide a starting point along
these lines. Secondly, our general expression for the nonforward impact 
factor will be of use also in electroweak physics where higher order
contributions to vector boson scattering are of interest ~\cite{peters}.

\bigskip
\centerline{\bf Acknowledgements}

We wish to thank M. Diehl and A. Mueller for very
helpful discussions.   K. P. is supported by the \textit{Graduiertenkolleg}
"Zuk\"unftige Entwicklungen in der Teilchenphysik". This research has been
supported in part by the Polish KBN grant No. 5 P03B 144 20 and the
Deutsche Forschungsgemeinschaft fellowship.
%%%%%%%%%%%%%%%%%%%%%%%%%%%%%%%%%%%%%%%%%%%%%%%%%%%%%%%%%%%%%%%%%%%%%%%%%%5
 
\end{document}

%% file: gaga.pstex_t
\begin{picture}(0,0)%
\includegraphics{gaga.pstex}%
\end{picture}%
\setlength{\unitlength}{4144sp}%
\begingroup\makeatletter\ifx\SetFigFont\undefined%
\gdef\SetFigFont#1#2#3#4#5{%
  \reset@font\fontsize{#1}{#2pt}%
  \fontfamily{#3}\fontseries{#4}\fontshape{#5}%
  \selectfont}%
\fi\endgroup%
\begin{picture}(5609,3128)(412,-3035)
\put(1844,-397){\makebox(0,0)[lb]{\smash{\SetFigFont{10}{12.0}{\familydefault}{\mddefault}{\updefault}{\color[rgb]{0,0,0}$u-q$}%
}}}
\put(432,-353){\makebox(0,0)[lb]{\smash{\SetFigFont{10}{12.0}{\familydefault}{\mddefault}{\updefault}{\color[rgb]{0,0,0}$q+l$}%
}}}
\put(2707,-353){\makebox(0,0)[lb]{\smash{\SetFigFont{10}{12.0}{\familydefault}{\mddefault}{\updefault}{\color[rgb]{0,0,0}$q-l$}%
}}}
\put(953,-1166){\makebox(0,0)[lb]{\smash{\SetFigFont{10}{12.0}{\familydefault}{\mddefault}{\updefault}{\color[rgb]{0,0,0}$k+l$}%
}}}
\put(2173,-1162){\makebox(0,0)[lb]{\smash{\SetFigFont{10}{12.0}{\familydefault}{\mddefault}{\updefault}{\color[rgb]{0,0,0}$k-l$}%
}}}
\end{picture}

%% file: paper82g.bbl
\newpage
\begin{thebibliography}{99}

\bibitem{Nikolaev}  N. N. Nikolaev and  B. G. Zakharov,
                   {\em Z. Phys.} \textbf{C 49} (1991) 607;
                   {\em Z. Phys}  \textbf{C 53} (1992) 331.

\bibitem{Mueller}   A. H. Mueller, {\em Nucl.\ Phys.} \textbf{B 415} (1994) 373;
                    {\em Nucl. Phys.} \textbf{B 437} (1995) 107.

\bibitem{GBW} K. Golec-Biernat and  M. W\"usthoff,
             {\em Phys. Rev.} \textbf{D 59}  (1999) 014017;
             {\em Phys. Rev.} \textbf{D 60} (1999)  114023;
             {\em  Eur. Phys. J.} \textbf{C 20} (2001) 313,

\bibitem{ForshawForw} J. R. Forshaw, G. Kerley and G. Shaw,
                 {\em Phys. Rev} {\bf D 60}  (1999) 074012,
                 {\em Nucl. Phys.} {\bf A 675} (2000) 80.

\bibitem{Levin}   E. Gotsman, E. M.  Levin, U. Maor and E. Naftali,
                 {\em Eur. Phys. J.} \textbf{C 10} (1999) 689.

\bibitem{McDermott}  M. McDermott, L. Frankfurt, V. Guzey and M. Strikman,
                   {\em Eur. Phys. J.} \textbf{C 16} (2000) 641.

\bibitem{BGBK}  J. Bartels, K. Golec-Biernat and  H. Kowalski,
              {\em Phys. Rev.} \textbf{D 66} (2002) 014001.

\bibitem{Cheng-Wu} H. Cheng and T. T. Wu,
              {\em Phys. Rev.} {\bf 182} (1969) 1852, 1868, 1873, 1899.

\bibitem{Lipatov}  L. N. Lipatov and G. V. Frolov,
                 {\em Yad. Fiz.} \textbf{13} (1971) 588.  


\bibitem{IK} D. Yu. Ivanov and R. Kirschner, {\em Phys.Rev.} {\bf D 58} (1998)
114026;
D. Yu. Ivanov, R. Kirschner, A. Schafer and  L. Szymanowski,
{\em Phys.Lett.} {\bf B 478} (2000) 101.

\bibitem{Wusthoff-Ivanov} D. Yu. Ivanov and M. Wusthoff
              {\em Eur.Phys.J.} {\bf C 8} (1999) 107.

\bibitem{ForshawNonForw} N. G. Evanson and J. R. Forshaw,
              {\em Phys. Rev.} {\bf D 60} (1999) 0340161.

\bibitem{Ryskin} M. G. Ryskin,
              {\em Z. Phys.} {\bf C 57} (1993) 89.


\bibitem{Gieseke-Qiao} S. Gieseke and C. F. Qiao,
              {\em Phys.Rev.} {\bf D 61} (2000) 074028.

\bibitem{Soper} S. J. Brodsky, F. Hauptmann and D. E. Soper,
              {\em Phys. Rev.} {\bf D 56} (1997) 6957.

\bibitem{kogut} J. B. Kogut and D. E. Soper, {\em Phys.Rev.} {\bf D 1} (1970) 2901.

\bibitem{Soper2} D. E. Soper, {\em Phys.Rev.} {\bf D 15} (1977) 1141.


\bibitem{Jeffbook}  J. R. Forshaw and  D. A. Ross,
               {\it Quantum Chromodynamics and the Pomeron},
               Cambridge University Press, Cambridge, England, 1996.

\bibitem{peters} K. Peters and G. P. Vacca, {\em in preparation}.

\end{thebibliography}
